\documentclass[aps,prb,twocolumn,superscriptaddress,groupedaddress,amsmath]{revtex4}  
\usepackage{graphicx}  
\usepackage{dcolumn}   
\usepackage{bm}        

\graphicspath {{figures/}}

\begin{document}

\title{Planar Hall-effect, Anomalous planar Hall-effect, and Magnetic Field-Induced Phase Transitions in TaAs}
\author{Q. R. Zhang}
\affiliation{National High Magnetic Field Laboratory, Florida State University, Tallahassee-FL 32310, USA}
\affiliation{Department of Physics, Florida State University, Tallahassee-FL 32306, USA}
\author{B. Zeng}
\affiliation{National High Magnetic Field Laboratory, Florida State University, Tallahassee-FL 32310, USA}
\author{Y. C. Chiu}
\affiliation{National High Magnetic Field Laboratory, Florida State University, Tallahassee-FL 32310, USA}
\affiliation{Department of Physics, Florida State University, Tallahassee-FL 32306, USA}
\author{R. Sch\"{o}nemann}
\affiliation{National High Magnetic Field Laboratory, Florida State University, Tallahassee-FL 32310, USA}
\author{S. Memaran}
\affiliation{National High Magnetic Field Laboratory, Florida State University, Tallahassee-FL 32310, USA}
\affiliation{Department of Physics, Florida State University, Tallahassee-FL 32306, USA}
\author{W. Zheng}
\affiliation{National High Magnetic Field Laboratory, Florida State University, Tallahassee-FL 32310, USA}
\affiliation{Department of Physics, Florida State University, Tallahassee-FL 32306, USA}
\author{D. Rhodes}
\affiliation{National High Magnetic Field Laboratory, Florida State University, Tallahassee-FL 32310, USA}
\affiliation{Department of Physics, Florida State University, Tallahassee-FL 32306, USA}
\author{K.-W. Chen}
\affiliation{National High Magnetic Field Laboratory, Florida State University, Tallahassee-FL 32310, USA}
\affiliation{Department of Physics, Florida State University, Tallahassee-FL 32306, USA}
\author{T. Besara}
\affiliation{National High Magnetic Field Laboratory, Florida State University, Tallahassee-FL 32310, USA}
\author{R. Sankar}
\affiliation{Center for Condensed Matter Sciences, National Taiwan University, Taipei 10617, Taiwan}
\author{ F. Chou}
\affiliation{Center for Condensed Matter Sciences, National Taiwan University, Taipei 10617, Taiwan}
\author{G. T. McCandless}
\affiliation{The University of Texas at Dallas, Department of Chemistry and Biochemistry, Richardson, TX 75080 USA}
\author{J. Y. Chan}
\affiliation{The University of Texas at Dallas, Department of Chemistry and Biochemistry, Richardson, TX 75080 USA}
\author{N. Alidoust}
\affiliation{Laboratory for Topological Quantum Matter (B7), Department of Physics, Princeton University, NJ 08544, USA}
\author{S.-Y. Xu}
\affiliation{Laboratory for Topological Quantum Matter (B7), Department of Physics, Princeton University, NJ 08544, USA}
\author{I. Belopolski}
\affiliation{Laboratory for Topological Quantum Matter (B7), Department of Physics, Princeton University, NJ 08544, USA}
\author{M. Z. Hasan}
\affiliation{Laboratory for Topological Quantum Matter (B7), Department of Physics, Princeton University, NJ 08544, USA}
\author{F. F. Balakirev}
\affiliation{National High Magnetic Field Laboratory, Los Alamos National Laboratory, MS E536, Los Alamos, New Mexico 87545, USA}
\author{L. Balicas}
\affiliation{National High Magnetic Field Laboratory, Florida State University, Tallahassee-FL 32310, USA}
\affiliation{Department of Physics, Florida State University, Tallahassee-FL 32306, USA}
\date{\today}

\begin{abstract}
We evaluate the topological character of TaAs through a detailed study of the angular, magnetic-field and temperature dependence of its
magnetoresistivity and Hall-effect(s), and of its bulk electronic structure through quantum oscillatory phenomena.
At low temperatures, and for fields perpendicular to the electrical current, we extract an extremely large Hall angle $\Theta_H$ at higher fields,
that is $\Theta_H \sim 82.5^{\circ}$, implying a very pronounced Hall signal superimposed into its magnetoresistivity.
For magnetic fields and electrical currents along the basal plane we observe a very
pronounced planar Hall-effect upon rotation of the field.
This effect, observed at temperatures as high as $T = 100$ K, is predicted 
to result from the chiral anomaly among Weyl points. Superimposed onto this planar Hall,
we also observe an anomalous planar Hall-signal akin to the one reported for ZrTe$_5$ and attributed to a non-trivial texture of its Berry phase.
Below 100 K, negative longitudinal magnetoresistivity (LMR), initially ascribed to the chiral anomaly and subsequently to current inhomogeneities,
is observed in all of the measured samples, once a large Hall signal is subtracted. 
We find that the small Fermi surface (FS) sheets of TaAs are affected by the Zeeman-effect as one approaches the quantum limit,
where a topological phase-transition is claimed to occur. This transition leads to
the reconstruction of the FS and to the concomitant suppression of the negative LMR
indicating that it is intrinsically associated with the Weyl dispersion.
For fields along the \emph{a}-axis it also leads to a pronounced hysteresis pointing to a field-induced electronic phase-transition.
This ensemble of unconventional tranport observations in TaAs points to a prominent role played by the axial anomaly among Weyl nodes
under the simultaneous application of magnetic and electric fields.
\end{abstract}

\maketitle

\section{introduction}
Weyl semi-metals, like the Ta and Nb based monopnictides, are characterized by  the lack of inversion, strong spin-orbit coupling, and
linearly dispersing bands in three dimensions which are modeled \emph{via} two copies of the Weyl equation \cite{ari,bernevig,hasan_Nat_Comm,jia}. The Weyl equation \cite{Weyl}
was originally proposed as an alternative to the Dirac one \cite{Dirac} for describing spin $1/2$ massless particles, or neutrinos. 
Band structure calculations\cite{bernevig,hasan_Nat_Comm} and angle resolved photoemission experiments,
find that these compounds are characterized by crossings between valence and conduction bands
at specific points in the Brillouin zone around which the bands disperse linearly as in Dirac systems\cite{hasan_Nat_Comm, ding_PRX, xu_science, xu_Nat_Phys, ding_Nat_Phys,yang_Nat_Phys}.
Theoretical calculations indicate that these so-called Weyl points occur in pairs, are topologically non-trivial, and act as either sinks or sources of Berry-phase
curvature, thus acting as topological charges \cite{bernevig,hasan_Nat_Comm}. Therefore, charge carriers undergoing electronic orbits around one, or in between a pair of these points,
would acquire a net chirality in their velocity and hence a concomitant Berry phase, thus effectively acting as Weyl fermions.
The calculations\cite{bernevig,hasan_Nat_Comm} predict the existence of 12 pairs of Weyl points in the first Brillouin zone of the Nb and Ta monopnitides, 4 pairs at $k_z = 0$ and 8 additional ones at $k_z \sim 0.59 \pi$.
Each pair is composed of Weyl points with opposite chirality or opposite topological charge. The projection of the band structure towards the surface of these materials leads
to topologically non-trivial surface states namely Fermi arcs connecting points on the surface which are the projection of the bulk Weyl
points \cite{bernevig, hasan_Nat_Comm, ding_PRX, xu_science, xu_Nat_Phys, ding_Nat_Phys,yang_Nat_Phys,Fermi_arcs}.
The separation between these points, and hence the length of the Fermi arcs, is found to increase with the strength of the spin-orbit coupling \cite{Fermi_arcs}.

The application of an external magnetic field is predicted to break the chiral symmetry between Weyl fermions \cite{adler,bell,nielsen,pallab,TaAs_QL,Zhang}.
This leads to the so-called Adler-Bell-Jackiw, or axial-anomaly \cite{adler,bell,nielsen,pallab,TaAs_QL,Zhang}, which corresponds to a net flow of Weyl fermions along
the axis connecting Weyl points of opposite chirality. For clean systems, and depending on the functional form of the quasiparticle scattering potential \cite{pallab},
this ``axial" current is predicted \cite{nielsen,pallab} to induce a net increase in the longitudinal magnetoconductivity (when $\mu_0 \overrightarrow{H} \| \overrightarrow{j}$,
where $j$ is the current density). This effect was claimed to have been observed in TaAs \cite{TaAs_QL,Zhang} although more recently it was argued to be an artifact resulting from
current inhomogeneity \cite{arnold1,liang}. The chiral anomaly is predicted to lead to novel effects, like a novel type
of quantum oscillatory phenomena in very thin crystals, i.e. cyclotron orbits traversing the bulk of the crystal (along the smallest dimension) and involving the Fermi arcs at the 
opposite surfaces of the crystal \cite{stern,moll} or to a planar Hall-effect for fields rotating within the plane of the electrical current \cite{burkov, tewari}. Elucidating the role of the chiral or axial anomaly
and exposing its related, novel optoelectronic properties is one of the most relevant current subjects in condensed matter physics and is the leading motivation for this study.

Since the FSs of semi-metallic systems, in particular those of the monopnictides, are rather small \cite{Zhang,moll,Moll,arnold1, arnold,lee,luo}, it is possible to reach
the quantum limit (QL) with available magnetic fields. In its conventional definition, the quantum limit is reached when the energy of all Landau levels, with the exception of
the $n=0$ level, exceeds that of the Fermi level according to the following dispersion relations:
\begin{align}\label{conventional}
  E(n,k_z) &= \left( n+\frac{1}{2}\right) \frac{\hbar eB}{\mu} + \frac{\hbar^2 k_z^{2}}{2\mu}\\
    &= \hbar v_F \sqrt{2B(n + \gamma)+ k_z^{2}}
\end{align}

Where  Eqs. (1) and (2) describe the electronic dispersions for conventional- and Weyl-like carriers in metallic systems, respectively. Here, $e$ corresponds
to the electron charge, $\hbar$ to the Planck constant, $v_F$ is the Fermi velocity, $\mu$  the carrier effective mass, $B$  the magnetic induction field,
and $\gamma$ the Berry phase of the charge carrier ($\gamma = 0$ for Weyl and Dirac systems). It turns out that a sharp anomaly, or an abrupt change in the sign of the magnetic torque,
was observed upon approaching the QL in NbAs\cite{Moll}. This observation was attributed to a change in regime, from a diamagnetic one intrinsic to the Weyl states below
the QL, to a paramagnetic one associated with states located above the Weyl node \cite{Moll}. There are also reports of anomalies in the longitudinal
magnetoresistivity of TaAs which are observed well beyond the QL for magnetic fields applied along its \emph{c}-axis \cite{Zhang2}.
These anomalies were suggested to result from electronic interactions, or a nesting instability among Weyl electrons possibly leading to helical spin-density waves\cite{Zhang2}.
Electronic instabilities, proposed to be magnetic field-induced charge-density waves associated with the quasi-one-dimensional electronic dispersion  described by Eq. (1),
are indeed observed in graphite beyond its QL \cite{fauque}. However, in elemental Bi, which as the monopnictides is characterized by the coexistence of
normal and relativistic carriers, the Nernst-effect at very high magnetic fields can be described within the non-interacting single particle picture \cite{zhu}.

Here, we evaluate in detail the magnetotransport properties of TaAs, magnetoresistivity and Hall-effect, as well as its bulk electronic structure at the Fermi level,
through magneto-transport measurements at low temperatures and at high magnetic fields surpassing its quantum limit. At $T = 100$ K its transverse magnetoresistivity is observed to increase nearly linear
in field by a factor $> 2000$ \% under $\mu_0H=9$ T. But for every sample measured, the magnetoresistivity is observed to decrease in amplitude upon further cooling
despite an expected increase in carrier mobility due to the suppression of phonon scattering. For every sample measured
we observe a pronounced assymetry between positive and negative fields due to a very pronounced, superimposed Hall signal. In contrast, the transverse Hall-effect increases
considerably upon cooling and leads to a very large Hall angle for a three-dimensional system, or
$\Theta_H \gtrsim 82.5^{\circ}$ at $T = 5$ K. For currents and fields along the basal plane, we
observe the emergence of a very large planar Hall signal as recently predicted for Weyl systems subjected to the Adler-Abel-Jackiw anomaly. 
Superimposed onto this planar Hall-effect, we also observe an anomalous planar Hall signal characterized by its assymetry with respect to positive and negative field values.
Below $\sim 50$ K and for all samples measured, we observe the emergence of negative longitudinal magnetoresistivity (LMR) once the superimposed Hall signal is subtracted. This
negative LMR survives until the quantum-limit is reached, where a topological/electronic phase-transition occurs.
For fields along the basal plane and the \emph{c}-axis, we observe a change in the oscillatory pattern of the Shubnikov de Haas oscillations upon surpassing the quantum limit,
pointing to Fermi surface reconstruction. We also observe hysteresis associated with this transition for fields aligned along the basal plane. 
This indicates that the longitudinal negative magnetoresistivity is associated to the original Weyl dispersion at the Fermi level. 
The observation of a very large Hall angle, a planar Hall-effect, an anomalous planar Hall signal, negative magnetoresistivity that is suppressed at the quantum limit, 
all indicate a non-trivial texture for the curvature of the Berry phase in TaAs. These observations support previous experimental results and theoretical predictions 
claiming a primordial role for the axial anomaly in this and related compounds. 

\section{Experimental}
\begin{figure*}[htp]
\begin{center}
    \includegraphics[width = 16 cm]{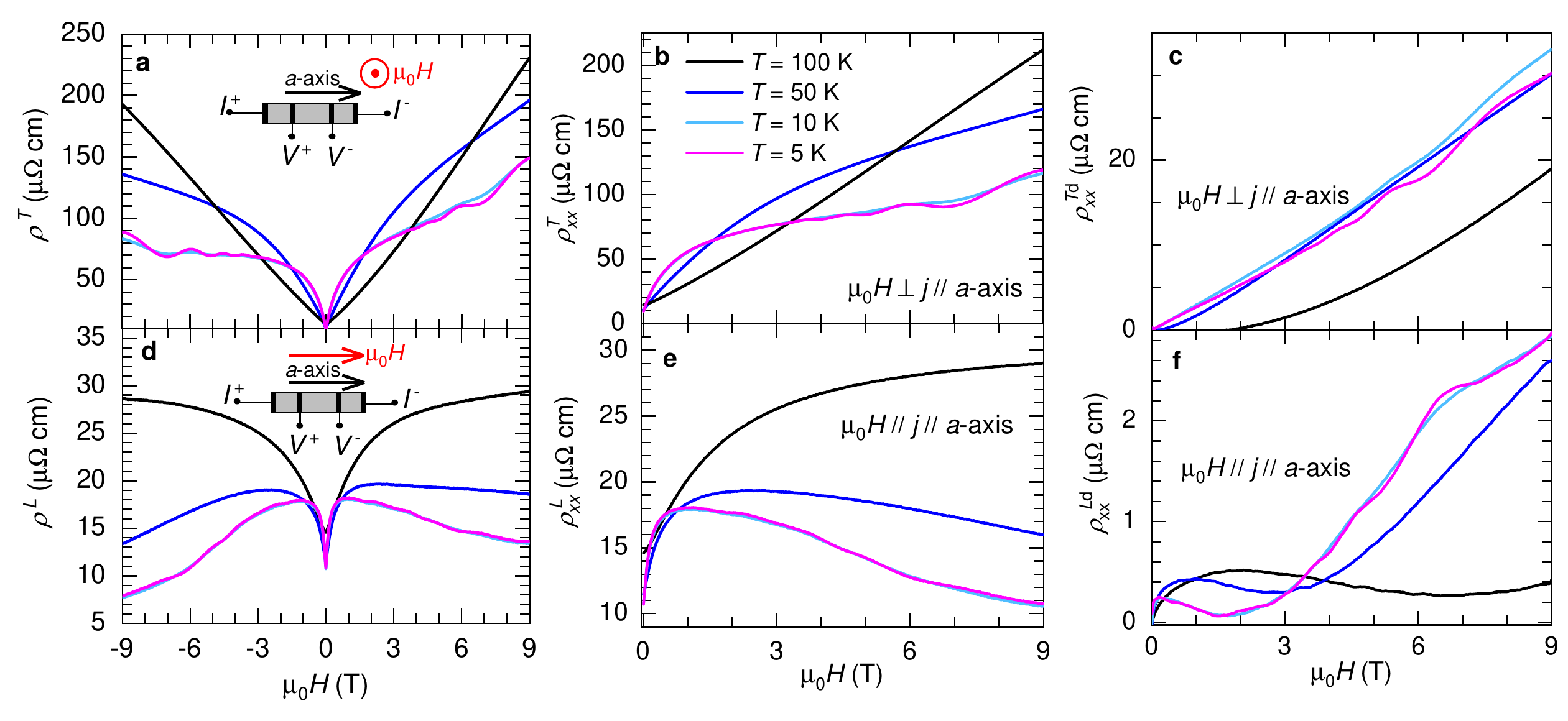}
    \caption{(a) Tranverse magnetoresistivity $\rho^T$ for a TaAs crystal (crystal \# 1)
    as a function of the field $\mu_0H$ applied along its \emph{c}-axis and for currents flowing along the \emph{a}-axis of the crystal and for several temperatures $T$.
    There is a marked assymetry in $\rho(\mu_0H)$ between positive and negative field values despite the use of a conventional four-terminal configuration for the electrical contacts.
    (b) Average tranverse magnetoresistivity $\rho^T_{xx}$ extracted from both positive and negative field traces. (c) Difference between both traces which yields a
    Hall-effect like signal $\rho^{T\text{d}}_{xx}(\mu_0H)$. (d) Longitudinal magnetoresistivity $\rho^L$ for the same TaAs crystal as a function of the
    field $\mu_0$H applied along the electrical current or along the \emph{a}-axis of the crystal and for several temperatures $T$. Notice the asymmetry with respect to field
    orientation as well as the decrease in $\rho^L$ as the field increases. (e) Average longitudinal magnetoresistivity $\rho^L_{xx}$ as function of $\mu_0H$.
    Below $T = 50$ K, and as $\mu_0H$ increases, $\rho^L_{xx}$ shows an initial increase which is followed by a pronounced decrease, as initially reported by Ref. \cite{chiral}.
    (f) Difference between positive and negative field $\rho^{L\text{d}}_{xx}$ traces, suggesting the superposition of an anomalous longitudinal Hall-like response. }
\end{center}
\end{figure*}

Single crystals of monoarsenides were grown \emph{via} a chemical vapor transport technique, as previously described \cite{besara}.
Polycrystalline precursor specimens were first prepared by sealing elemental Ta and As mixtures under vacuum in quartz
ampoules and by heating the mixtures at a rate of 100 $^{\circ}$C/h to 700 $^{\circ}$C, followed by a dwell for 3 days. The
polycrystalline TaAs boules were subsequently sealed under vacuum in quartz ampoules with iodine to serve as the transporting agent.
The ampoules were slowly heated up in a horizontal tube furnace under a temperature gradient $\Delta T$ = 100 $^{\circ}$C.
The ampoules were maintained under this condition for 3 weeks before rapid cooling them down to room temperature.
This process produced a large number of single crystals  with typical dimensions of 0.5 mm$^3$.

The crystallographic axes of the measured single-crystals were identified through single crystal X-ray diffraction
and subsequently polished to produce bar shaped crystals with their main axis aligned along a specific crystallographic orientation.
Torque magnetometry was measured using a CuBe cantilever beam technique whose deflection was measured capacitively.
Electrical contacts were produced by attaching gold wires with silver paint in a standard four-terminal configuration
for either magnetoresistivity or Hall effect measurements. Transport measurements were performed \emph{via} a conventional
AC technique in either a Physical Properties Measurement System, a 35 T resistive magnet, or the 45 T hybrid magnet at the NHMFL in Tallahassee in combination
with either a $^3$He cryostat or variable temperature insert. Measurements were also performed in a 60 T pulse field magnet at the NHMFL-LANL (not included here).
\begin{figure*}[htp]
\begin{center}
    \includegraphics[width = 16 cm]{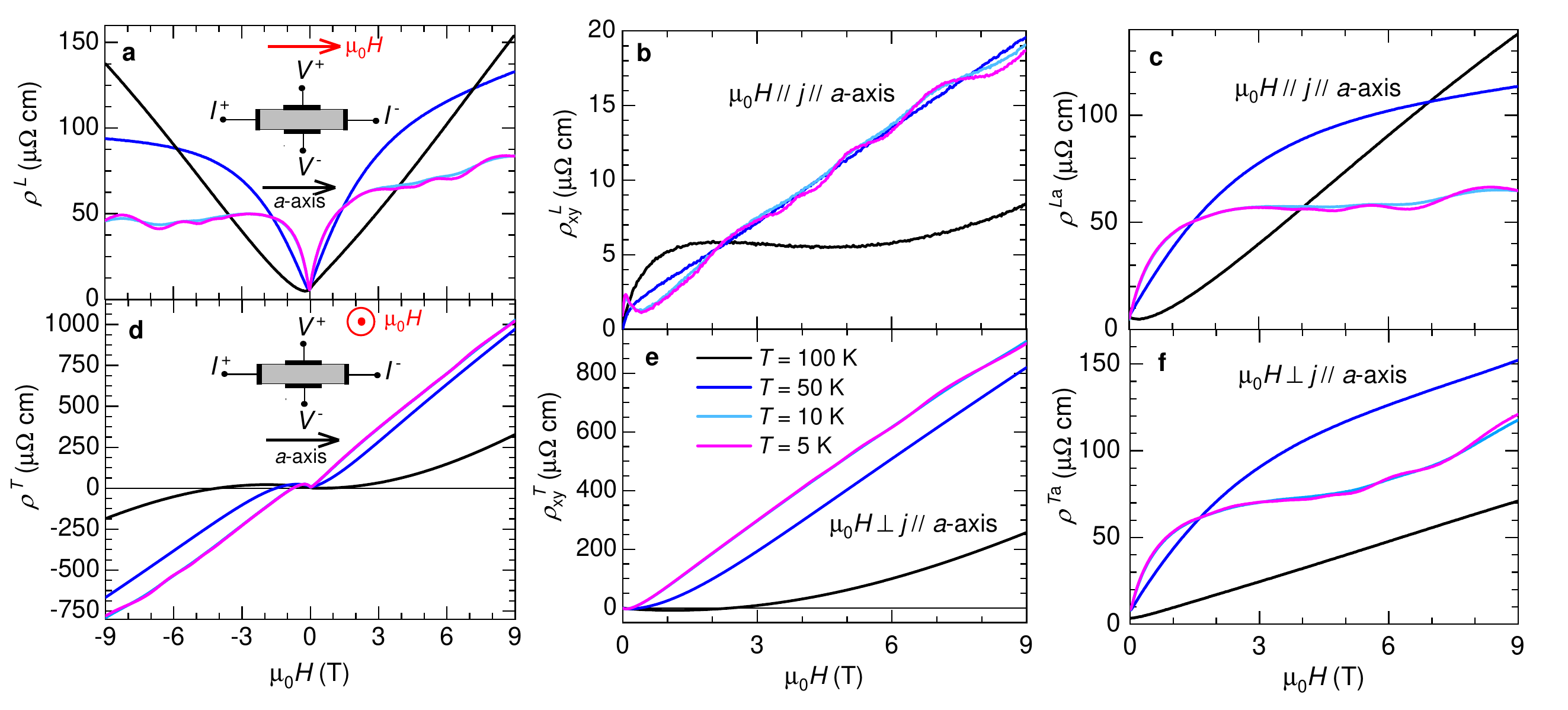}
    \caption{(a) Longitudinal resistivity $\rho^L$ using a Hall configuration of contacts for currents and fields parallel to the
    \emph{a}-axis, as collected from the TaAs crystal \# 1 under several temperatures. (b), Difference between traces collected under positive and negative fields or
    $\rho_{xy}^L$. Below $T = 50$ K it displays a nearly linear dependence in field . (c) Average between both traces or $\rho^{L\text{a}}$, which yields the superimposed
    longitudinal magnetoresistivity. (d) Raw Hall-effect traces, or $\rho^T$, measured also from crystal \# 1 under fields along its \emph{c}-axis. (e)
    Transverse Hall-effect or $\rho^T_{xy}$ resulting from the difference between both traces after multiplication of the raw signal by the sample thickness $t$. (f) Average
    between both traces or the superimposed transverse magnetoresistivity signal $\rho^{T\text{a}}$.}
\end{center}
\end{figure*}

Figure 1 provides an overall evaluation of the magnetoresistivity $\rho(H)$, for fields parallel (longitudinal) and perpendicular (transverse) to the electrical current
density $\overrightarrow{j}$ which flows along the basal plane of a TaAs crystal. See Supplemental Information file for a picture of the crystal in Fig. S1 \cite{supplemental}. 
This crystal, as well as all the other crystals studied here, was oriented through single crystal X-ray
diffraction and polished to reduce its thickness along the \emph{c}-axis resulting, in this case, in a relatively thick platelet having a thickness $t \simeq 0.3$ mm. Crystals
measured at high fields were also polished to similar thicknesses. To minimize the role of current inhomogeneities, we chose a conventional four-terminal configuration as depicted
in the sketch within Fig. 1(a) that shows the transverse magnetoresistivity $\rho^T$ as a function of the magnetic field $\mu_0H$, for both positive and negative values,
and for several temperatures. Notice i) the asymmetry between positive and negative fields despite the use of a non-Hall configuration of contacts,
ii) the nearly linear in field magnetoresistivity observed at $T= 100$ K and beyond, and iii) its decrease as the temperature is lowered followed by saturation
as the field increases. This last observation is striking given the suppression of phonon scattering as \emph{T} is lowered which should increase the carrier lifetime
and lead to a concomitant increase in magnetoresistivity. Figure 1(b) plots the average $\rho_{xx}^T$ between positive and negative field traces showing that $\partial \rho_{xx}^T/\partial(\mu_0H)>0$
over the entire temperature range. In contrast, Fig. 1(c) plots the difference $\rho_{xx}^{Td}$ between positive and negative field traces revealing a nearly linear in field
or a Hall-like signal superimposed onto the magnetoresistivity. Below, the reader will find measurements of the transverse Hall-effect measured on the same sample using
a Hall configuration of electrical contacts and revealing strikingly similar traces at these temperatures. Figure 1(d) plots the longitudinal magnetoresistivity $\rho^L$, that is
for $\mu_0H \| j \|a$-axis, as a function of $\mu_0H$ and for several $T$s. There is a very small assymetry between positive and negative traces collected at $T=100$ K. The saturation 
of $\rho^L$ at higher fields, is akin to the behavior reported many years ago by Pippard for conventional metals \cite{pippard}. But as \emph{T} is lowered, one
observes instead the emergence of negative magnetoresistivity, i.e. $\partial \rho_{xx}^L/\partial(\mu_0H)<0$, after an initial increase of $\rho^L(\mu_0H)$, as previously
reported in Ref. \onlinecite{chiral}. Figure 1(e) displays the average $\rho^L_{xx}$ between these positive and negative field traces, indicating that at lower $T$s $\rho^L_{xx}$
decreases by over 40 \% between $\mu_0H =1$ and 9 T. Figure 1(f), displays their difference or $\rho_{xx}^{Ld}$ which, as we shall see in Fig. 2, present strong similarities with
traces collected using a Hall-effect like configuration of electrical contacts intended for Hall-effect measurements due to, as we shall see, an anomalous planar Hall-effect.

\begin{figure}[htp]
\begin{center}
    \includegraphics[width = 8.6 cm]{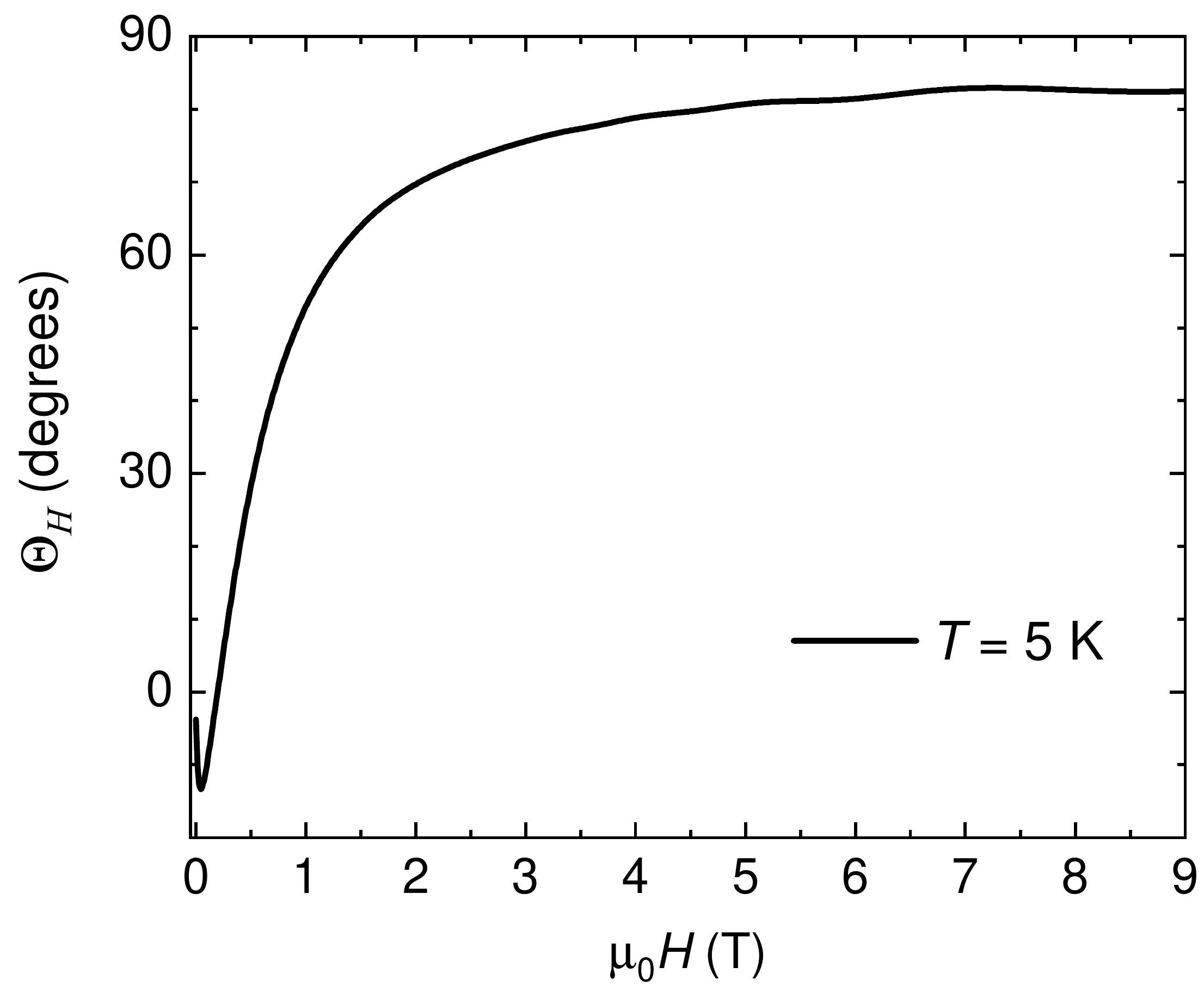}
    \caption{Hall angle $\Theta_H = \arctan (E_y/E_x) = \rho_{xy}^T/\rho_{xx}^T $ as a function of $\mu_0H$, as extracted
    from Figs. 1(b) and 2(e), for $T = 5$ K.  At high fields the Hall angle saturates at a rather large value for a three dimensional system, i.e.
    $\Theta_H(\mu_0H >> 1) \simeq 82.5^{\circ}$.  }
\end{center}
\end{figure}
\begin{figure*}[htp]
\begin{center}
    \includegraphics[width = 18 cm]{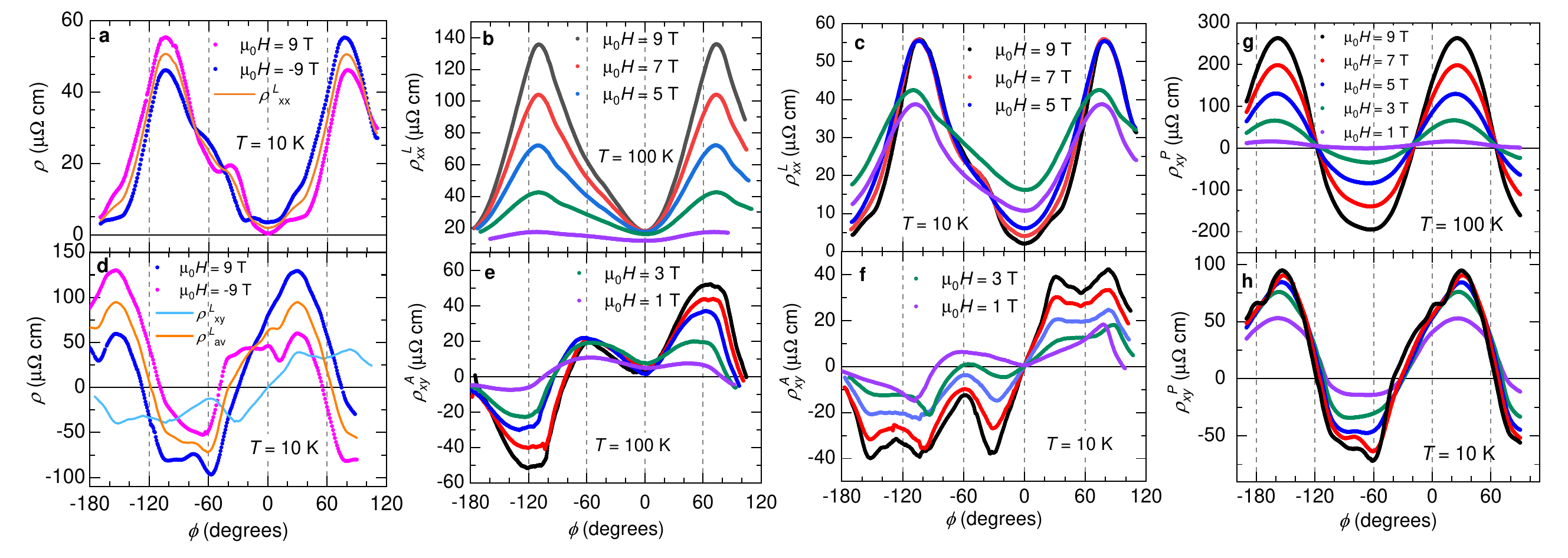}
    \caption{(a) Angular dependence of the magnetoresistivity $\rho$ (conventional configuration of contacts), at
    a temperature $T = 10$ K and for $\mu_0H = 9$ T (magenta markers) and -9 T (blue markers). $\phi$ is the planar angle between the electrical current injected along the
    \emph{a}-axis and the external field. The angular asymmetry between both traces contrasts with the symmetry shown by their average $\rho_{xx}^L$ (orange markers)
    which displays a periodicity of $180^{\circ}$. (b) $\rho_{xx}^L$ as function of $\phi$ at $T = 100$ K and for several fields. For all angles, $\rho_{xx}^L$ increases
    as the field increases. (c), $\rho_{xx}^L$ as function of $\phi$ at $T = 10$ K and for several fields. In contrast, $\rho_{xx}^L (T = 10 \text{ K})$ decreases as
    $\mu_0H$ increases, except around a narrow angular range around its maxima where it is observed to increase. For all three panels a conventional four-terminal configuration of
    contacts was used. (d) Angular dependence of the longitudinal resistivity measured at $T = 10$ K through a Hall configuration of contacts and for both field orientations.
    Here, clear orange markers depict the average between both traces or $\rho_{xy}^P$, while the blue ones depict their difference or the assymmetric component $\rho_{xy}^L$.
    (e) Anomalous planar Hall $\rho_{xy}^L$ as a function $\phi$ at $T = 100$ K and for several field values. (f) $\rho_{xy}^L$ as a function $\phi$ at $T = 10$ K and for several
    field values. (g) Planar Hall response $\rho_{xy}^P$ for fields rotating within planes at \emph{T} = 100 K and for
    several fields. (h) $\rho_{xy}^P$ at $T = 10$ K and under a few values of the external field. Panels (b), (c),
    (e), (f), (g) and (h) use the same color scheme to identify the applied magnetic field value, namely black, red, blue, green and violet markers
    depict $\mu_0 H = 9$, 7, 5, 3 and 1 T, respectively.}
\end{center}
\end{figure*}

\begin{figure*}[htp]
\begin{center}
    \includegraphics[width = 18 cm]{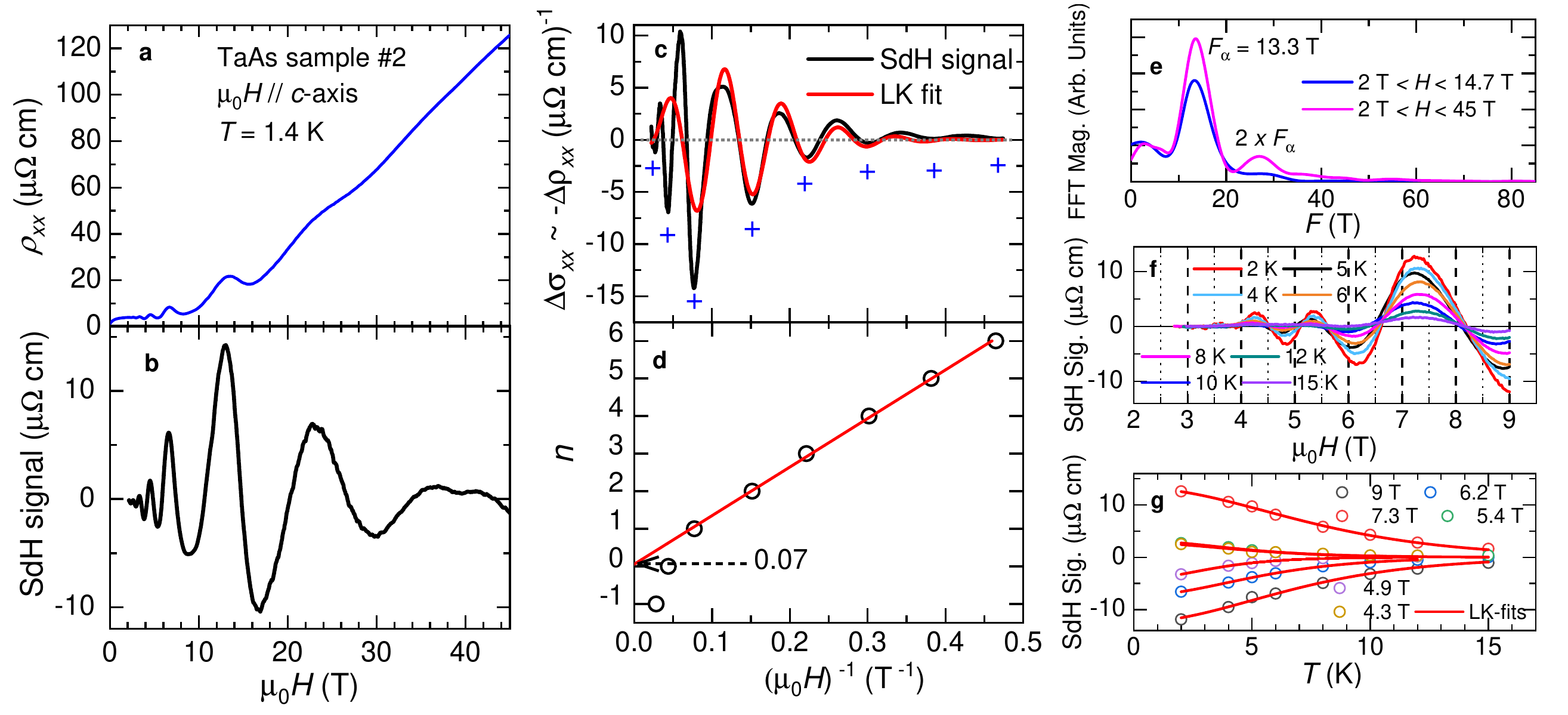}
    \caption{(a) Magnetoresistivity of a second TaAs single-crystal (sample \# 2)
    at $T = 1.4$ K and as function of the field $\mu_0H$ applied along its \emph{c}-axis.  (b) SdH signal $\Delta \rho_{xx}$ superimposed onto the magnetoresistivity after
    background subtraction. (c) Oscillatory signal in the conductivity $\Delta \sigma_{xx} \sim - \Delta \rho_{xx}$ as a function of $(\mu_0H)^{-1}$. Red line is a fit to a single
    Lifshitz-Kosevich oscillatory term. This single oscillatory component cannot reproduce the data at low fields, nor at high fields indicating a Fermi surface
    that is field-dependent. The abrupt change in periodicity observed above $\mu_0H \sim 16$ T cannot be attributed to magnetic breakdown due to the disappearance of the
    periodicity observed at lower fields indicating instead Fermi surface reconstruction. Crosses indicate the positions of the minima in $\Delta \sigma_{xx}$. The fit yields
    $\Phi \simeq (2\pi \times -0.091)$ corresponding to $\phi_B = 2 \pi \times (0.53 \pm 0.03)$. However, restricting the fit to $\mu_0H < 5$ T yields
    $\Phi \simeq (2 \pi \times 0.62)$ due to the Zeeman-effect induced deformation of this FS pocket. (d) Landau index plot as a function $(\mu_0H)^{-1}$.
    The minima observed at the highest fields do not follow the same Landau index sequence. Red line is a linear fit yielding $\Phi \simeq (2 \pi \times 0.07)$ or
    $\phi_B \sim (2 \pi \times 0.7)$. (e) FFT of $\Delta \rho_{xx}$ taken at two field intervals revealing a single prominent peak at $F_{\alpha} = 13.3$ T, and also its second
    harmonic when data at the highest fields are included. (f) Oscillatory signal superimposed on the Hall-effect of sample \# 2 as a function of $\mu_0H$ and for several temperatures.  (g)
    Amplitude of the maxima and minima observed in the oscillatory signal as a function of $T$. Red lines are fits to the LK formula which yield effective masses
    between 0.15 and 0.2 $m_0$.}
    \label{TaAs_Summ1}
\end{center}
\end{figure*}

Figure 2 provides an overall evaluation of the Hall-effect, measured on the same crystal for fields along both the basal plane and its \emph{c}-axis.
For these measurements, we carefully removed the original voltage leads initially configured for collecting the magnetoresistance traces. To minimize the role of current
inhomogeneities or contact misalignment with respect to the equipotential lines we chose, not point-like, but wide Hall contacts convering a sizeable portion along the longitudinal
dimension of the sample, as depicted in the inset of Fig. 2(a). Given the Hall configuration, we multiplied all traces in Fig. 2 by the only dimensional factor
relevant to the Hall-effect, namely the sample thickness $t \simeq 0.03$ cm. As we just mentioned the observation of a planar Hall-effect, we choose to initially show
the raw longitudinal resistivity $\rho^L$ measured when  $j \| \mu_0H$. Figure 2 (a) displays a plot of $\rho^L$ as a function of $\mu_0H$, showing an anomalous, nearly linear 
and non-saturating LMR at $T = 100$ K, which contrasts with the conventional, saturating LMR seen at lower $T$s. Remarkably, no negative LMR is observed when using this configuration of contacts,
although it is in principle more sensitive to inhomogenous current distributions \cite{ZrTe5}. This fact points to a relatively uniform current density. Figure 2(a) also reveals 
asymmetry between positive and negative field traces whose difference $\rho_{xy}^L$ is displayed in Fig. 2(b). $\rho_{xy}^L$ displays a nearly linear in field behavior, with all
curves collected below $T= 50$ K overlapping one another, as one would expect for a Hall-like signal. Figure 2(c) displays the average between both traces $\rho^{La}$, which
corresponds to the LMR signal superimposed onto this anomalous Hall signal, revealing just conventional saturating LMR. Figure 2(d) displays the raw transverse or conventional 
Hall-effect $\rho_{xy}^T$ measured for $j \bot \mu_0H \| c$-axis. Figure 2(e) shows the difference between traces collected under positive and negative fields and thus corresponds 
to the pure transverse Hall-effect $\rho_{xy}^T$. It displays a distinct functional form (with respect to field) as well as a different temperature dependence when compared to
$\rho_{xy}^L$, thus indicating that $\rho_{xy}^L$ cannot be attributed to field misalignment with respect to the planar direction. The very large values reached by
$\rho_{xy}^T$ when compared, for instance, to $\rho{xx}^T$ in Fig. 1(a) points to a large Hall angle. Finally, Fig. 2(f) displays the magnetoresistivity component
$\rho^{Ta}$ superimposed onto $\rho_{xy}^T$, or the average between the positive and the negative field traces in Fig. 2(d). As expected, this component is rather small
relative to the Hall signal.

Given the very large Hall signal seen in Fig. 2, subsequently, we evaluate the Hall angle $\Theta_H = \arctan \left( E_y/E_x \right) =  \arctan \left( \mu_0H t R_H/ \rho_{xx}^T \right) = \arctan \left( \rho_{xy}^T/ \rho_{xx}^T \right)$.
Figure 3 displays $\Theta_H = \arctan \left( \rho_{xy}^T/ \rho_{xx}^T \right)$ for $T = 5$ K as a function of $\mu_0H$ indicating that $\Theta_H$ saturates at a very large value
$\Theta_H \simeq 82.5^{\circ}$ for a three-dimensional system. Such a large Hall angle, which is relatively close to $\Theta_H \simeq 90^{\circ}$ observed in two-dimensional
quantum Hall-effect (QHE) systems, implies that it is very difficult to collect magnetoresistivity signal without having to subtract the superimposed Hall-effect,
regardless of the configuration of electrical contacts or the geometry of the sample, as is the case for QHE systems.
Current jetting, or a magnetic field-dependent current density distribution within single-crystals of monoarsenides, was recently claimed to be responsible for
the observation of the negative LMR \cite{arnold1,liang} initially ascribed to the axial anomaly \cite{chiral}. However, given the very
large $\Theta_H$, or the superposition of a large Hall component that could easily be misinterpreted as evidence for inhomogeneous current distribution, throughout the remainder
of this manuscript we will present the raw magnetoresistivity data as well as the average between positive and negative field traces to subtract the aforementioned component.

In Fig. 4 we present a detailed characterization of the angular dependence of $\rho(\mu_0H, \phi)$ for $\mu_0H(\phi) \bot c$-axis, using both the magnetoresistive and
the Hall configuration of electrical contacts previously used in Figs. 1 and 2, respectively. Here, $\phi$ is the angle between $\overrightarrow{j}$ and $\mu_0\overrightarrow{H}$
applied along the basal plane, where $\phi = 0^{\circ}$ corresponds to $\mu_0H \| j$. Figure 4(a) shows the raw $\rho (\phi)$ traces collected at $T=10$ K for $\mu_0H = +9$ T
(magenta markers) and -9 T (blue makers), when using the conventional four-terminal configuration for magnetoresistivity measurements. There is a clear asymmetry between positive
and negative field traces due to the superimposed Hall component. Orange markers depict the average $\rho_{xx}^L$ between both traces which reveals maxima having the same amplitude
at either side of the minima. Figure 4(b) plots $\rho_{xx}^L$ under several fields at $T = 100 $ K as a function of $\phi$, where it displays an unexpected saw tooth-like dependence
instead of a simple sinusoidal as one would expect for a system driven from zero to maximum Lorentz force upon rotation. At this temperature, $\rho(\mu_0H)$ remains
always positive and increases with field regardless of its orientation. In contrast, Fig. 4(c) displays $\rho_{xx}^L$ as a function of $\phi$ for several fields but at $T = 10 $ K
revealing a broad range of angles, at either side of the minima, where $\rho_{xx}^L$ decreases upon increasing the field. As in Figs. 1 and 2, the maximum value of
the magnetoresistivity decreases when $T$ is lowered. The sharp difference between panels 4(b) and 4(c), namely the observation of negative LMR, is difficult to reconcile with
the possibility of current jetting dominating the transport at low $T$s particularly when contrasted with Fig. 2 where it is absent.

Figure 4(d) displays $\rho(\phi, \mu_0H = \pm 9 \text{T})$ for the same crystal but obtained with the Hall configuration of contacts used to extract the data in Fig. 2.
A marked asymmetry is again observed between the trace collected under $\mu_0H = +9$ T (blue markers) and the one collected under -9 T (magenta). Orange markers depict their average,
which is an odd function of the angle (in contrast to the traces in panels 4(b) and 4(c) and akin to the so-called planar Hall-effect, observed recently in GdPtBi \cite{felser}
and Cd$_3$As$_2$ \cite{planar_Hall_Cd3As2}, and which is claimed to result from the axial anomaly between Weyl points \cite{burkov,tewari}. The planar Hall-effect is usually
observed in ferromagnetic metals or semiconductors subjected to strong spin-orbit coupling and interpreted as resulting from the anisotropy in the magnetoresistivity induced
by the inherent magnetic anisotropy of the system. But in non-magnetic Weyl systems the chiral anomaly is predicted to induce a giant planar Hall-effect (PHE), or the
appearance of a large transverse voltage leading to $\rho_{xy} \propto \sin(2\phi)$\cite{burkov,tewari}. For this experimental configuration one should not observe any Hall-like signal under normal circumstances, in
contrast to what is indicated by the orange markers in Fig. 4(d). Remarkably, the asymmetry between positive and negative field traces leads to another signal distinct from the PHE,
also odd in angle, and as a true Hall-effect it is an odd function of the field (clear blue markers) and therefore akin to the anomalous planar Hall-effect (APHE) recently
reported \cite{ZrTe5} for ZrTe$_5$. According to the authors of Ref. \onlinecite{ZrTe5}, the close correlation between the temperatures where the negative LMR
and the anomalous planar Hall-effect are observed, indicates that the Weyl nodes and related Berry-phase texture are responsible for this APHE. Figures 4(e) and 4(f) display this
anomalous planar Hall signal $\rho_{xy}^A$ for several field values and for two temperatures $T =$ 100 K and 10 K, respectively. Notice that i) $\rho_{xy}^A$ is observed at a
temperature ($T = 100$ K) where one does \emph{not} observe negative LMR, in contrast to Ref. \onlinecite{ZrTe5}, and ii) its maximum value as a function of the
angle remains nearly constant as $T$ is lowered. At the moment, we do not have an adequate theoretical framework to explain this effect,
to understand its precise functional form or its evolution as a function of $T$. Nevertheless, the traces collected at $T=10$ K are likely to be affected
by the angular dependence of the superimposed Shubnikov-de-Haas-effect which is discussed below. Figs. 4g and 4h plot the PHE signal for several field values and respectively,
for $T = 100$ and 10 K. As expected, and in contrast to $\rho_{xx}^L(\phi)$ in Fig. 4(b), at $T = 100 $ K the PHE signal is nearly sinusoidal with its maxima displaced by
$\sim 45^{\circ}$ with respect to the maxima in $\rho_{xx}^L(\phi)$, as previously reported for both GdPtBi \cite{felser} and Cd$_3$As$_2$ \cite{planar_Hall_Cd3As2}.
Its periodicity, i.e. $\sim 180^{\circ}$, would indicate Weyl cones displaying no tilting with respect to the Fermi level \cite{tewari}. The non-sinusoidal nature
of the PHE at $T = 10$ K clearly results from the superposition of the SdH signal. Finally, the magnitude of both the PHE and the APHE in TaAs is considerably smaller
than the respective values reported for Cd$_3$As$_2$, GdBiPt and ZrTe$_5$.
\begin{figure*}[htp]
  \centering
    \includegraphics[width = 18 cm]{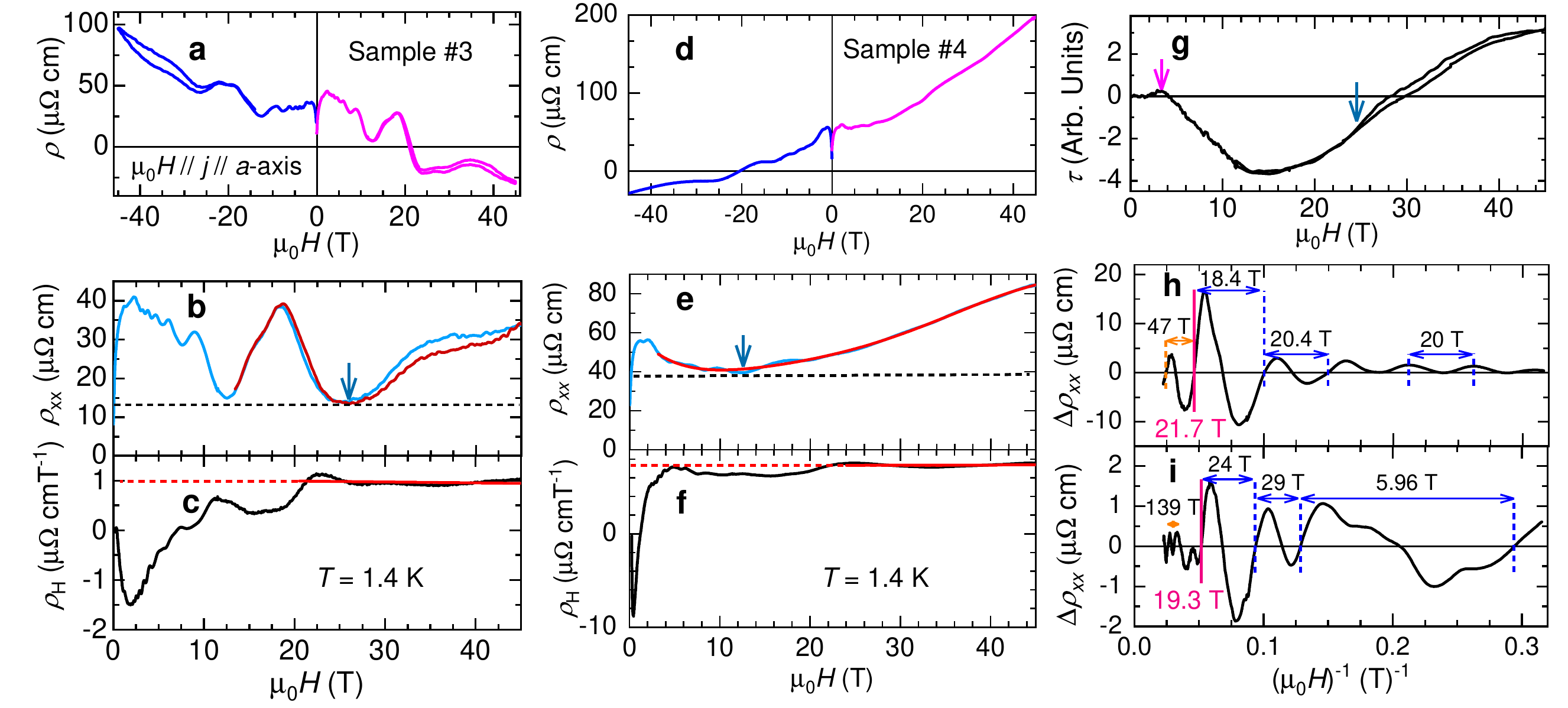}
    \caption{(a) Raw resistivity $\rho$ for a second TaAs single-crystal
    (sample \# 3) as a function of $\mu_0H \| a$-axis at $T = 1.4$ K. $\rho (\mu_0H)$ displays a marked asymmetry with respect to both field orientations, indicating the superposition
    of a Hall-like component. (b) Averaged resistivity, or $\rho_{xx} = (\rho(\mu_0 H)+ \rho(-\mu_0 H))/2$ as a function of $\mu_0H$. $\rho_{xx}$ decreases beyond
    $\mu_0H \sim 2$ T as previously observed and claimed to result from the axial current among Weyl points. (c) Assymmetric or Hall-like component
    $\rho_H = (\rho(\mu_0 H)- \rho(-\mu_0 H))t/(2\mu_0H) $ as a function of $\mu_0H$ where $t$ is the sample thickness. $\rho_H$ is non-linear but saturates beyond $\mu_0 H \sim 25$ T which is the field beyond which $\rho_{xx}$ increases
    again as function of $\mu_0H$ while also displaying hysteresis. Both observations point towards an electronic phase-transition which probably is also topological in character.
    (d) $\rho$ as a function of $\mu_0H$ for a fourth single-crystal (sample \# 4) also at $T = 1.4$ K. This sample displays a more pronounced assymmetric
    magnetoresistivity with respect to sample \# 3 indicating a larger  misalignment of $\mu_0H$ with respect to the $a$-axis. (e) and (f)
    $\rho_{xx}$ and $\rho_H$ for sample \# 4  as functions of $\mu_0H$, where the red line (in (e)) is an example of a polynomial fit to the magnetoresistive background.
    The oscillatory signal $\Delta \rho_{xx}$, superimposed onto $\rho_{xx}$, is obtained by subtracting the polynomial background from the experimental data. (g) Magnetic torque
    $\mathbf{\tau} = \bf{M} \times \mu_0 \bf{H}$ as a function of $\mu_0 \bf{H}$ oriented nearly along the $a$-axis under $T = 1.4$ K.
    Magenta arrow indicates an anomaly close to the value in field where the magnetoresistivity crosses from positive $(\partial \rho_{xx}/\partial(\mu_0H) >0)$ to
    negative $(\partial \rho_{xx}/\partial(\mu_0H) <0)$ as the field increases. Blue arrow indicates the onset of hysteresis. The overall behavior of $\tau$, i.e. from negative to positive values upon crossing the quantum limit, is akin to
    the observations of Ref. \cite{Moll}. (h) and (i) $\Delta \rho_{xx}$ as a function of $(\mu_0 H)^{-1}$ for samples \# 3 and \# 4, respectively.
    The difference in the periods of the oscillatory signals indicates a small difference in the relative position of the
    Fermi level(s) between both samples. The abrupt change in periodicity beyond a certain critical field is indicated by a pink vertical line.}
    \label{TaP_Summ2}
\end{figure*}

Figure 5(a) displays the transverse magnetoresistivity $\rho_{xx}$ for a second TaAs single-crystal, measured under fields up to $\mu_0H = 45$ T applied along its \emph{c}-axis,
at a temperature of $T=1.4$ K. Notice the pronounced magnetoresistivity, i.e. $\Delta \rho = (\rho (\mu_0H = 45 \text{T})-\rho (0 \text{T}))/\rho (0 \text{T}) \simeq 10^4$ \%,
the oscillatory signal, or the SdH-effect, superimposed onto $\rho(\mu_0H)$, and the sharp change in the slope of $\rho(\mu_0H)$ observed above $\mu_0H \sim 16$ T. This pronounced
increase coupled to the suppression of the amplitude of oscillatory signal and change in its frequency, points to the onset of the quantum limit. For this sample we extracted
an approximate Hall mobility of $\sim \mu_H \simeq 5.4 \times 10^5$ cm$^2$/Vs which is comparable to reported values for NbAs \cite{ghimire} being about one order of magnitude
higher than those reported for TaP \cite{arnold}. Nevertheless, the correct extraction of the Hall mobilities of these compounds would require a two-band analysis of their
non-linear Hall response \cite{arnold1}. Figures 5(b) and 5(c) display the oscillatory component $\Delta \rho_{xx}$, or the Shubnikov de Haas signal, superimposed onto the
$\rho_{xx}(\mu_0H)$ as function of $\mu_0H \parallel c-$axis and of its inverse, respectively. To obtain the SdH signal, the resistivity as function of the field was fit to a
polynomial which was subsequently subtracted from it. The oscillatory signal changes its amplitude and periodicity upon approaching the quantum limit:
its amplitude grows as a function of the magnetic field up to $\mu_0H_{QL} \sim 13$ T, which is the value where one observes
a sharp increase in $\rho_{xx}(\mu_0H)$. Beyond this point the oscillations not only decrease in amplitude but, as seen in Fig. 5(c), their period also decreases.
This becomes clear from a comparison between the experimental data and a fit to a single Lifshitz-Kosevich (LK) oscillatory component (red line).
The oscillatory component in the magnetoconductivity $\Delta \sigma_{xx} \sim -\Delta \rho_{xx}$ as a function of $(\mu_0 H)^{-1}$ can
be described by the Lifshitz-Kosevich formalism \cite{Shoenberg}:
\begin{eqnarray*}
\Delta \sigma[(B)^{-1}] \propto \frac{T}{B^{5/2}}\sum_{l=1}^{\infty} \frac{\exp^{-\pi l/ \omega_c \tau}\cos(l g \mu \pi/2)}{l^{3/2}\sinh(\alpha T/B)}
\end{eqnarray*}
\begin{eqnarray}
\times \cos\left\{ 2\pi \left[ \left( \frac{F}{B}-\frac{1}{2}+\phi_B \right)l \pm \delta \right] \right\}
\end{eqnarray}
where $F$ is the dHvA frequency, $l$ is the harmonic index, $\omega_c$ the cyclotron frequency, $g$ the Land\'{e} \emph{g}-factor, $\mu$ the effective mass in
units of the free electron mass $m_0$, and $\alpha = 14.69$ is a constant. $\delta$ is a phase-shift determined by the dimensionality of the Fermi surface acquiring values
of either $\delta = 0$ or $ \pm 1/8$ for two- and three-dimensional Fermi surfaces, respectively.  $\phi_B$ is the Berry phase which, for Dirac and Weyl systems
\cite{bernevig}, is predicted to acquire $\phi_B = \pi $. Therefore, its precise value defines the topological nature of any given compound.
From the fit one extracts a Berry phase of $\phi_B = 2\pi \times (0.53 \pm 0.03)$, or a non-trivial Berry phase in contrast to Ref. \onlinecite{chiral} which finds
$\phi_B \simeq 1/8$. However, the fit is far from excellent probably due to a Zeeman-effect induced evolution of the geometry of the Fermi surface.
In fact, if the field window is limited to $\mu_0 H \leq 5$ T the fit yields $\phi_B \sim 0$, indicating that the Zeeman-effect indeed precludes the extraction
of the Berry-phase of TaAs through quantum oscillations. In Fig. 5(c) the blue crosses indicate the minima in the SdH signal, used to
create the Landau plot shown in Fig. 5(d), which are associated with the frequency $F \simeq 13$ T dominating the fast Fourier transform (FFT) spectra shown in Fig. 5(e).
If the topography of the Fermi surface of TaAs was field independent, the quantum limit (or the $n=0$ Landau level) should be reached at fields just above
$\mu_0H_{QL} \simeq F_{\alpha} \sim 13$ T. The role of the quantum limit becomes apparent in Fig. 5(d) which displays the Landau index $n$ of the peaks observed
in $\Delta \rho_{xx}$ as a function of the their positions in $(\mu_0H)^{-1}$, with the red line being a linear fit whose intercept yields $\phi_B \simeq 0.7$, again distinct
from the previously extracted values. The minima observed in $\Delta\rho_{xx}(\mu_0H)$ above $\mu_0H \sim 16$ T deviate from the original sequence indicating either Fermi surface
reconstruction around the quantum limit or the emergence of fractionalized excitations \cite{yu}. Figure 5(e) shows the FFT spectra for two field intervals, i.e. $2 \text{ T} \leq \mu_0H \leq 14.7 \text{ T}$
(blue line) and $2 \text{ T} \leq \mu_0H \leq 45 \text{ T}$ (magenta line). One observes a main peak at $F_{\alpha} \simeq 13$ T, while the inclusion of the oscillatory signal
observed at the highest fields leads to the emergence of a second peak at a frequency that one would associate with its second harmonic $\sim 2F_{\alpha}$. Below, we show results
for other TaAs single-crystals indicating that the periodicity observed beyond the quantum limit would be sample dependent or rely on the precise position of the Fermi level.
Therefore, we conclude that the few oscillations observed by us in the different crystals beyond $\mu_0H_{QL}$ ought to be associated with an electronic reconstruction at
the Fermi level.

Figure 5(f) displays the SdH signal superimposed on the Hall-effect measured in sample \#1. We selected the Hall signal due to its near linearity as a function of the field which leads
to a reliable background subtraction. Figure 2(g) displays the amplitude of the peaks and valleys observed in Fig. 2(f) as a function of $T$.
Red lines are fits to the LK formalism from which we extract the effective mass $\mu$ (in units of $m_0$). These fits yield consistent values ranging between 0.15 and 0.2 $m_0$
even for the oscillations observed at the highest fields.

Given the previously described large Hall components, i.e. transverse Hall, PHE, and APHE superimposed on the magnetoresistivity of TaAs and their interplay with the
negative LMR observed below $T=50$ K, it is pertinent to ask how reproducible this last effect is among samples having different geometries and configurations
of electrical contacts. It is also relevant to evaluate how this effect can be influenced by the reconstruction of the Fermi level observed in the neighborhood
of the quantum limit. In Fig. 6 we evaluate the LMR, that is $\rho_{xx}(\mu_0H)$ for $\mu_0\bm{H} \| \bm{j} \| a-$axis, for other two polished TaAs single crystals (samples \#3 and \#4)
having thicknesses approaching 100 $\mu$m, using two point-like voltage contacts at one side of these crystals, see Ref. \onlinecite{supplemental}. This configuration leads to a stronger mixing between
Hall and magnetoresistivity signals. As mentioned in the Introduction, the application of a component of the magnetic field along an applied electric field is
predicted to break the chiral symmetry among Weyl points, creating a net flow of charge carriers, or an axial current, from one Weyl point to the other having
opposite chirality \cite{adler,bell,nielsen,pallab}. This axial current is predicted to induce a negative longitudinal magnetoresistivity indeed
observed in TaAs\cite{chiral,Zhang}, TaP \cite{chiralTaP}, Cd$_3$As$_2$ \cite{Cd3As2}, Na$_3$Bi \cite{Na3Bi}, and ZrTe$_5$ \cite{ZrTe5_chiral}. In contrast, the large negative
LMR observed in TaP was claimed to result from a field-induced inhomogeneous current distribution \cite{arnold1}, suggesting that this so-called current jetting effect
might have contaminated the results in those reports \cite{chiral,Zhang,chiralTaP, Cd3As2,Na3Bi,ZrTe5_chiral}. A more recent transport study found that current jetting
would be irrelevant to Na$_3$Bi or GdBiPt but would dominate the longitudinal transport in monoarsenides \cite{liang}.

To address this issue, in Fig. 6(a) we evaluate the longitudinal transport of sample \#3 under fields all the way up to $\pm 45$ T. Here, great care was taken
to align the sample along the basal plane with use of a Hall sensor and through the overall behavior of the sample (e.g. \emph{via} minimizing the Hall component on the data).
At these high fields very small misalignments, by a fraction of a degree, lead to large transverse Hall and transverse magnetoresistive components on the data.
One observes a marked asymmetry between positive (magenta trace) and negative (blue) field sweeps, although these traces were collected from both contacts situated on the same
side of the sample with the intention of evaluating its LMR. Undoubtedly, this asymmetry results from the large superimposed Hall component: a mixture of the APHE
and the transverse Hall-effect resulting from the lack of a high precision in the angular positioning $(\leq 0.3^{\circ})$.  Therefore, the average between both traces
should yield the pure LMR component. As seen in Fig. 6(b), the LMR of sample \#3 and for $\mu_0H\|a-$axis displays 3
regions: i) a low-field region dominated by positive magnetoresistivity observed in all previous studies \cite{arnold1,chiral,Zhang,chiralTaP,Cd3As2,Na3Bi,ZrTe5_chiral} and claimed to
result from three-dimensional weak anti-localization \cite{Zhang}, ii) a region characterized by \emph{negative} LMR claimed to result from the axial anomaly,
and iii) a high field region, whose onset is indicated by the vertical blue arrow, that is characterized by positive and hysteretic magnetoresistivity
pointing to a first-order phase-transition once the quantum limit is surpassed. The negative LMR is observed from higher field values than those in Ref. \onlinecite{Zhang} likely
due to the misalignment between $\mu_0H$ and the $a-$axis. Figure 6(c) plots the Hall component normalized by the field indicating that a well-defined Hall-constant is observed
only beyond this transition. Panels 6(d), 6(e0, and 6(f) present similar data for sample \#4. The larger values attained by its raw magnetoresistivity, when compared to sample \#3, indicate
a poorer sample alignment. This is also evident from the narrower range in fields where the negative LMR is observed or the larger values of the Hall component (i.e.
a sizeable transverse Hall signal). Therefore, at low temperatures all three crystals display a region in field where the LMR presents a negative slope, once the
Hall component(s) are subtracted, despite the differences in sample dimensions and in their configurations of contacts.
\begin{figure}[htp]
  \centering
    \includegraphics[width = 8.6 cm]{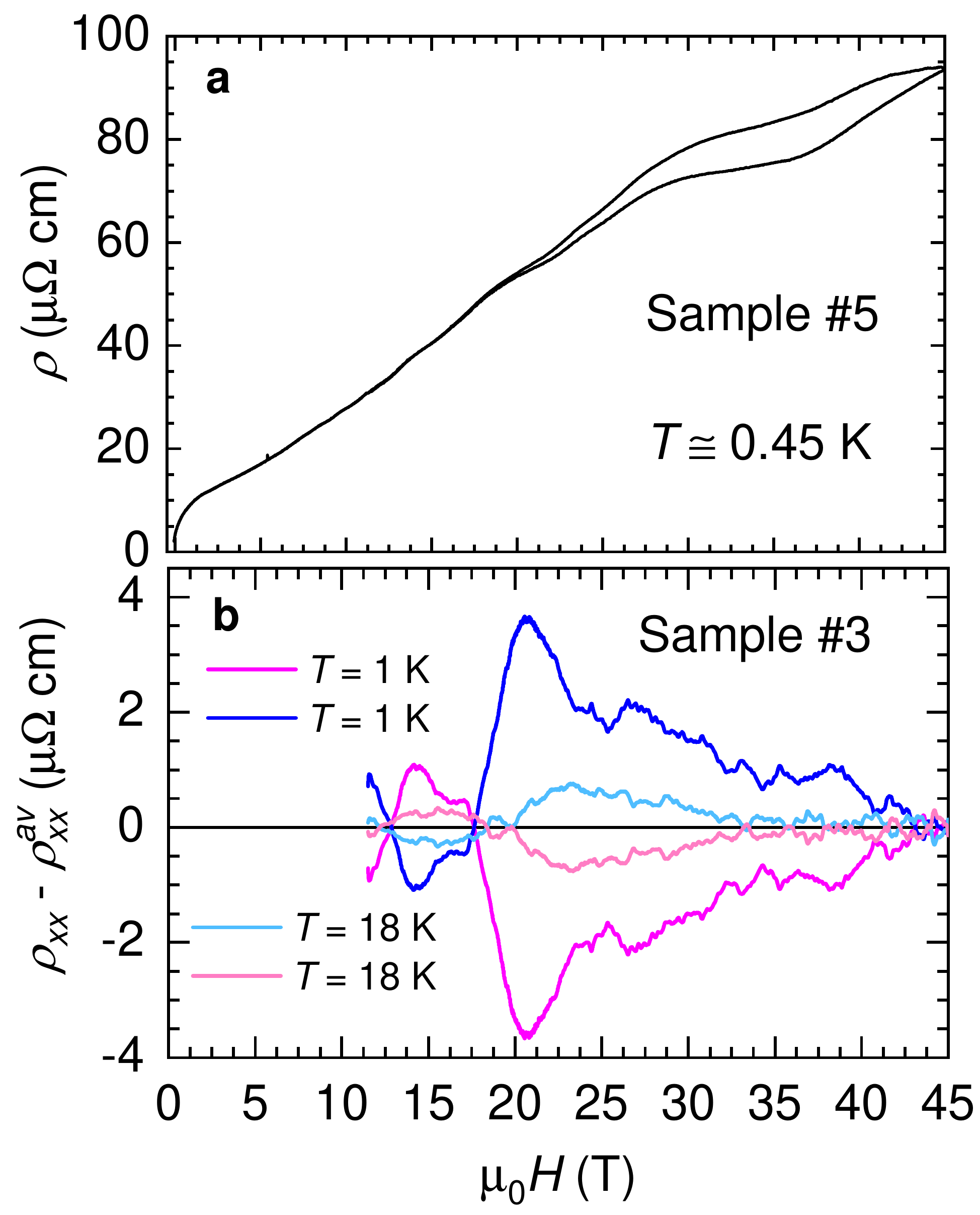}
    \caption{(a) $\rho(\mu_0 H)$ for a fifth TaAs single-crystal (sample \# 5) at an angle $\theta \sim 1.5^{\circ}$ with respect to the \emph{a}-axis. A pronounced hysteresis 
    observed beyond $\mu_0 H \simeq 20$ T, i.e. $\Delta \rho \sim$ 20 $\mu \Omega$ cm at $\mu_0=35$ T. (b) $\rho_{xx}- \rho^{av}_{xx}$ for sample \# 3 and for
    increasing (magenta lines) and decreasing (blue) field sweeps for two temperatures, $T = 1$ K and $T=18$ K. $\rho^{av}_{xx}$ is the average value between increasing and decreasing sweeps. }
    \label{hysteresis}
\end{figure}
\begin{figure}[htp]
\begin{center}
\includegraphics[width = 8 cm]{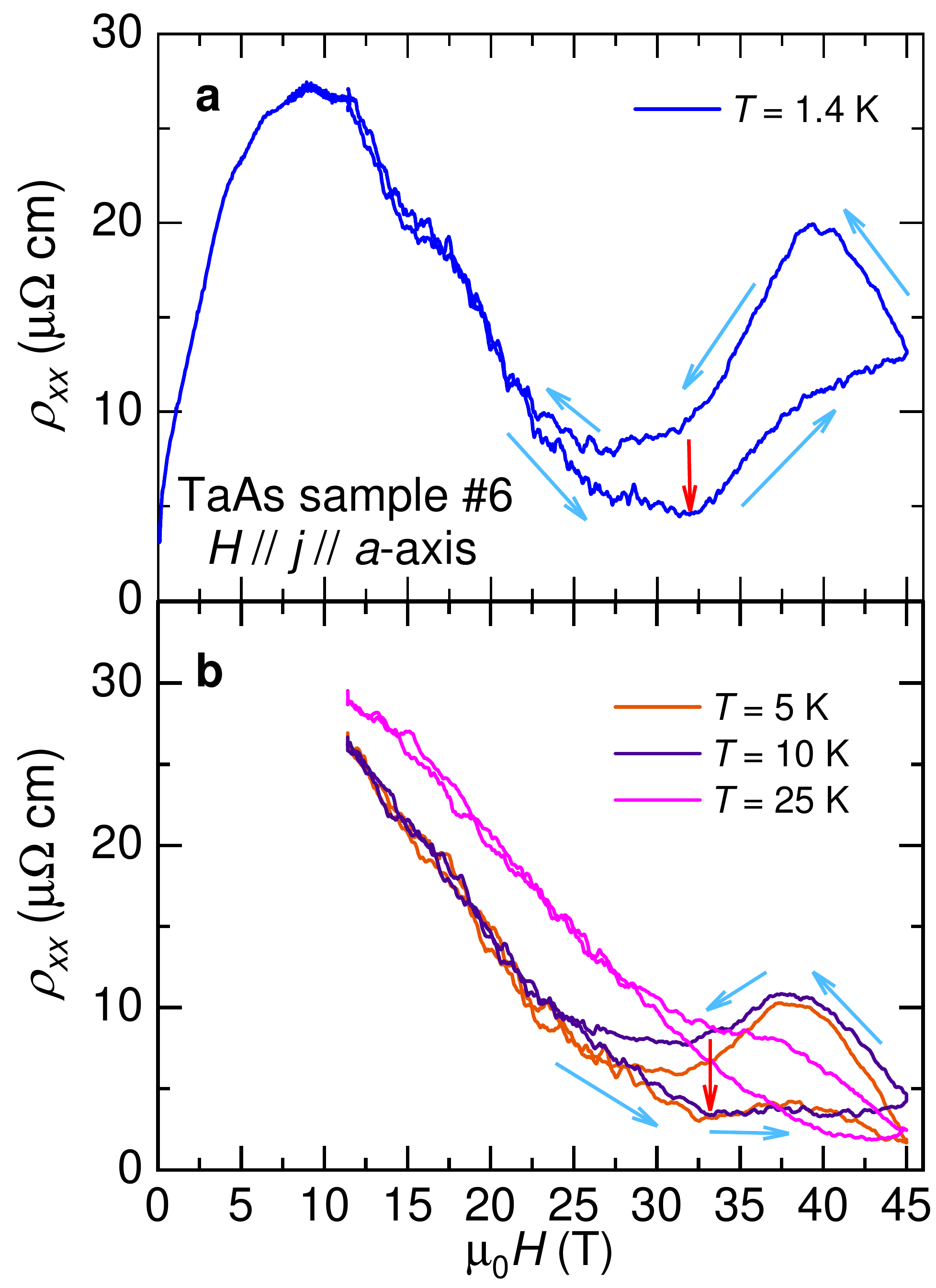}
\caption{(a) Non-symmetrized $\rho_{xx}$ for a sixth TaAs crystal as a function of $\mu_0 H$ applied nearly along the electrical current $j$ injected along the $a-$axis of the
crystal at a temperature $T = 1.4$ K. Its initial positive slope becomes negative beyond $H \sim 8$ T, an effect ascribed to the axial anomaly
    among Weyl points. For fields above $H_c \simeq 32.5$ T (indicated by the red arrow), the negative slope becomes positive again producing a large
    hysteresis as the field is swept to lower values. This change in slope combined with the hysteresis indicates a magnetic field-induced first-order phase transition.
    $H_c$ is close in value to the frequency $F$ of the superimposed weak oscillatory signal, or $F = (30 \pm 5)$ T indicating that it occurs in the neighborhood of the quantum limit.
    (b) $\rho_{xx}$ as a function of $H \parallel a-$axis for several temperatures. The hysteresis survives under temperatures as high as 25 K.}
    \label{TaP_Summ}
\end{center}
\end{figure}

To provide thermodynamic evidence for a hysteretic phase-transition beyond the QL, we measured the magnetic torque
$\overrightarrow{\tau} = \mu_0 \overrightarrow{M} \times \overrightarrow{H}$ of yet another TaAs single-crystal at $T = 1.4$ K, and as a function of $\mu_0H$ along a direction tilted
a few degrees with respect to the basal plane. Here, $\overrightarrow{M}$ is the sample's magnetization. $\tau$ as a function of $\mu_0 H$ is shown in Fig. 6(g)
revealing the behavior already described in the Introduction, namely a negative response at low fields, associated with the diamagnetism inherent to Weyl dispersion, which
switches to a positive one as the QL is reached and as previously reported for NbAs \cite{Moll}. The magnetic torque also reveals mild hysteresis emerging at nearly the
same field value, i.e. $\mu_0H \sim 25$ T, where the negative LMR in sample \#3 becomes suppressed. Figures 6(h) and 6(i) plot the oscillatory signal $\Delta \rho_{xx}$ superimposed onto
the LMR of samples \#3 and \#4, respectively. In these panels we included the frequencies associated to the observed oscillations. The difference in the oscillatory patterns
between both samples is due to a small difference in the relative position of their chemical potentials. There is an abrupt change in the sample dependent oscillatory pattern 
once the external field surpasses a value in the neighborhood of 20 T which is slightly higher than the value found for fields along the \emph{c}-axis. Despite the limitations imposed by a
maximum field of 45 T, both Figs. 5 and 6 indicate that the oscillatory pattern changes abruptly when the QL limit is surpassed indicating Fermi surface reconstruction and not
the opening of a gap everywhere within the Brillouin zone. Recently, an electronic phase-transition was observed in TaAs for fields (in excess of $\mu_0H=60$ T) and currents
applied along its \emph{c}-axis\cite{brad}. It is characterized by a $\sim 2$ orders of magnitude increase in magnetoresistivity and claimed to result from the opening of a gap
at the Weyl nodes \cite{brad}. Our results, namely the abrupt change in the quantum oscillatory pattern, the suppression of negative LMR for fields along the \emph{a}-axis, and
the observation of hysteresis in transport and in thermodynamic quantities, indicate that the Weyl dispersion is affected by the application of much lower fields that in reality do not
gap the entire Fermi surface. Although the observation of hysteresis does indicate a first-order phase-transition towards a field-induced phase akin to those observed
in graphite \cite{fauque}, or in quasi-one-dimensional organic systems \cite{FISDW1, FISDW2} which partially gaps their Fermi surface(s). It might correspond to
a magnetic field-induced charge-density wave state having for its modulation the wave-vector connecting both Weyl nodes of opposite chiralities as predicted
for pyrochlore iridates \cite{iridates}. The field-induced opening of a gap was indeed predicted for the monoarsenides based on \emph{ab-initio} calculations
due to the mixing of the zeroth Landau levels associated with the Weyl points of opposite-chirality \cite{chiral_breakdown}. This effect would be rather anisotropic,
occurring at much lower fields when these are aligned along the \emph{a}-axis. This contrasts with our observations indicating electronic reconstruction of the Fermi surface
for both field orientations at considerably lower fields than those reported in Refs. \onlinecite{brad} and \onlinecite{chiral_breakdown}. In Supplemental Fig. S2 \cite{supplemental},
we include Hall-effect measurements up to $\sim 63$ T applied along the $c-$axis of another TaAs single-crystal, indicating a decrease in the slope of the Hall resistivity 
$\rho_{xy}(\mu_0H)$ upon surpassing the quantum limit. This could be interpreted as either i) an increase in carrier density which would be at odds with the disappearance of
the small frequencies seen in the Shubnikov-de Haas signal or ii) the loss of the high mobility Weyl electrons originally located at the Fermi level. Naively, 
for a high mobility semimetal like TaAs, one would expect the loss of carriers to lead to poorer carrier compensation and hence to a higher Hall constant.
The opposite is observed, therefore we conclude that this transition leads to an overall decrease in carrier mobility due to the gapping of the Fermi surfaces located on the Weyl 
dispersing bands. This Hall signal, collected under pulsed fields, also shows another change in slope around $\mu_0H \simeq 50$ T which coincides with the transition
reported by Ref. \onlinecite{brad}. If the first change in slope is attributable to the loss of one set of Weyl electrons (e.g. W1), this second transition would be attributable 
to either the loss of the second set of Weyl carriers (at W2 points) or of the conventional holes (see, Fig. 1(a) in Ref. \onlinecite{brad}), in contrast to what is claimed by 
Ref. \onlinecite{brad}. Our results present certain similarities with those of Ref. \onlinecite{weyl_annihilation} which in TaP observes a series of anomalies in the magnetotransport 
properties below and above the quantum limit where the Hall-effect is observed to change sign abruptly. This was attributed to the gapping of the Weyl nodes resulting from the 
competition between the magnetic wave-vector and the separation in \emph{k}-space of the W1 nodes.

Figure 7(a) plots the non-symmetrized magnetoresistivity of a fifth crystal sample \#5 whose electrical current is not as well aligned with respect to the field as the previous crystals.
Therefore, the slope of its magnetoresistivity is always positive. One observes a pronounced hysteretic response emerging at higher fields and at lower temperatures ($T = 0.45$ K).
Figure 7(b) displays the hysteretic component or $\Delta \rho_{xx}(\mu_0H)=\rho_{xx}(\mu_0H)-\rho_{xx}^{av}(\mu_0H)$, where $\rho_{xx}^{av}(\mu_0H)$ is the average between increasing
and decreasing field sweep traces. These traces correspond to sample \#3 with the magenta and blue lines depicting increasing and decreasing field sweeps, respectively. The
hysteresis is observed well below the QL (lower fields are limited by the background field, or 11.5 T, of the superconducting coil of the hybrid magnet) and is still present at
much higher temperatures.  

Finally, Fig. 8(a) displays the non-symmetrized $\rho_{xx}(\mu_0H)$ for $\mu_0\bm{H}$ nearly along $\bm{j} \| a-$axis, and for yet a sixth TaAs single-crystal.
Blue arrows indicate both field increasing and decreasing sweeps.  The SdH oscillations observed in the background of the magnetoresistivity yields a frequency $F \sim (30 \pm 5 )$ T
which essentially coincides with $H_{\text{kink}}$ (indicated by red the arrow), where the slope of the magnetoresistivity changes from negative to positive and
which, combined with the pronounced hysteresis, indicates a first-order transition upon reaching the quantum limit. Figure 8(b) displays the longitudinal
magnetoresistivity as a function of $\mu_0H$ for several temperatures. Notice the disappearance of the sharp kink at $H_{\text{kink}}$ and of the concomitant
hysteresis upon increasing the temperature which indicates that this phase-transition emerges upon decreasing $T$. Therefore, Figs. 6 and 8, or samples \#3 and \#6,
indicate quite clearly that the negative LMR observed in TaAs correlates with its zero-magnetic field electronic structure at the Fermi level. TaAs undergoes
a first-order electronic phase-transition upon approaching the quantum limit which seemingly gaps the Weyl nodes and leads to a reconstructed Fermi surface. According to
Ref. \onlinecite{brad}, this reconstructed FS undergoes yet additional field-induced phase-transitions at higher fields.

\section{Summary}
The ensemble of anomalous transport properties shown here, namely an anomalously large Hall angle for a three-dimensional system, the observation
of a large planar Hall-effect and of an anomalous planar Hall-effect in this non-magnetic compound, and of a negative slope for the longitudinal
magnetoresisivity emerging only below $\sim 50$ K, points to the unique texture for the Berry-phase and to the prominent role played
by the axial/chiral anomaly in TaAs. Negative longitudinal magnetoresistivity \cite{chiral,Zhang,chiralTaP, Cd3As2,Na3Bi,ZrTe5_chiral,liang} and a large
planar Hall-effect have been predicted theoretically \cite{burkov,tewari} and observed experimentally in several compounds \cite{felser,planar_Hall_Cd3As2}.
Although the anomalous planar Hall-effect has just been reported for ZrTe$_5$ \cite{ZrTe5}, both effects are
observed simultaneously in TaAs requiring a theoretical effort to clarify their origin and interplay. From an experimental perspective,
it would be important to clarify if both effects disappear under fields beyond the quantum limit which seemingly gap the Weyl nodes.
This would unambiguously attach the nodes to their observation.

The observation of a hysteretic phase-transition confirms that the electronic structure of TaAs undergoes an electronic
and hence topological phase-transition(s) upon approaching and surpassing the quantum limit. The suppression of the negative longitudinal magnetoresistivity,
theoretically attributed to the axial anomaly, necessarily points to the gapping of the Weyl points. Its observation in multiple samples,
having different geometries and configuration of the electrical contacts \cite{supplemental} makes it nearly impossible to attribute negative longitudinal magnetoresistivity
to current inhomogeneities \cite{arnold1}, especially when combined with the observation of planar Hall and anomalous planar Hall-effects.
This co-called current jetting would not explain the re-emergence of positive magnetoresistivity at the hysteretic phase-transition
observed at the highest magnetic fields. Notice that the authors in Ref. \onlinecite{arnold1} did not consider the prominent role played by the Hall-effect(s) as done here,
which is very likely to influence their observations and conclusions. We conclude that the negative longitudinal magnetoresistivity observed in the monoarsenides is intrinsically
associated to their unique, Weyl-like electronic dispersion at the Fermi level.

Finally, the phase-transition observed here, in Ref. \onlinecite{brad}, and in TaP, upon approaching and surpassing the quantum limit are likely to result
from a competition between the magnetic wavelength and the wave-vector(s) connecting Weyl points of opposite chirality  \cite{weyl_annihilation}.
This competition is akin to what is observed in bulk quasi-one-dimensional organic conductors when undergoing a
cascade of magnetic field-induced spin-density wave transitions displaying concomitant quantum Hall-effect \cite{FISDW1,FISDW2}.
We suggest the Hall-effect to be carefully investigated under the highest fields in monopinictide samples having well-defined geometries.

\section{Acknowledgements}
L.~B. is supported by DOE-BES through award DE-SC0002613.
Work at Princeton University was supported by the
Gordon and Betty Moore Foundations Emergent
Phenomena in Quantum Systems Initiative through Grant
No. GBMF4547 (M. Z. H.). F.~C. acknowledges the support provided by
MOST-Taiwan under project no. 102-2119-M-002-004.
J.~Y.~C. acknowledges support from NSF DMR-1700030.
The NHMFL acknowledges NSF support through DMR-1157490
and DMR-1644779, and the State of Florida.

\end{document}